\documentstyle[rotating,11pt,newpasp,twoside,epsf,psfig,mathtime]{article}

\newcommand{\degree}{\mbox{$^{\circ}$}}               
\newcommand{\oversim}[2]{\protect{\mbox{\lower0.5ex\vbox{%
   \baselineskip=0pt\lineskip=0.2ex
   \ialign{$\mathsurround=0pt #1\hfil##\hfil$\crcr#2\crcr\sim\crcr}}}}} 
\newcommand{\etal}{\mbox{\hbox{\it et\,al.}}}         
\newcommand{\eg}{\mbox{\hbox{\it e.g.,}}}             
\newcommand{\thetaonec}{\mbox{$\theta ^1$Ori\,C}}     
\newcommand{\thetaoned}{\mbox{$\theta ^1$Ori\,D}}     
\begin{document}
\title{Wide-Field Thermal Imaging of the Orion Nebula \\
       at High Spatial Resolution}
\author{Lynne K. Deutsch, Marc Kassis, Nathan Smith\altaffilmark{1}}
\affil{Astronomy Department, Boston University, \\
       725 Commonwealth Avenue, Boston, MA 02215 \\
       deutschl, mkassis, nathans@bu.edu}

\author{Joseph L. Hora, Giovanni G. Fazio}
\affil{Smithsonian Astrophysical Observatory, \\
60 Garden Street MS/65, Cambridge, MA 02138, USA \\
jhora, gfazio@cfa.harvard.edu}

\author{Harold M. Butner, William F. Hoffmann}
\affil{Steward Observatory, University of Arizona, \\
933 North Cherry Avenue, Tucson, AZ 85721, USA \\
hbutner, whoffman@as.arizona.edu}

\author{Aditya Dayal}
\affil{Infrared Processing and Analysis Center, \\
MS 100-22, Caltech, Pasadena, CA 91125, USA \\
adayal@ipac.caltech.edu}

\altaffiltext{1}{Present address: Dept.\ of Astronomy, University of
Minnesota, 116 Church St.\ SE, Minneapolis, MN\,55455, USA; 
nathans@astro.umn.edu}

\begin{abstract}
We describe multi-wavelength (8--20\micron), diffraction-limited, mid-infrared 
images of the OMC-1 cloud core in Orion, covering an approximately two 
arcminute area around the Trapezium and BN/KL regions. We have detected
mid-infrared emission at the locations of a 
subset of the previously identified proplyds in the Orion Nebula along with 
two new infrared sources. The Ney-Allen nebula surrounding the OB star
\thetaoned{} exhibits a ring or toroidal structure at the longest wavelengths.  
The BN/KL complex appears as an extended, butterfly-shaped structure with
significant bipolar symmetry which is bifurcated by a dust lane at the
longer wavelengths. The infrared sources IRc3,~4, and~5 give the appearance 
of a ring-like structure with a possible jet-like protrusion from its center 
along a line from IRc2. Derived color temperature and dust opacity maps 
suggest that IRc3,~4, and~5 may not be self-luminous objects.
\end{abstract}
\section{Introduction}
The OMC-1 cloud core in Orion contains two areas of active and recent star 
formation, namely the Trapezium Cluster in the Orion Nebula and the embedded 
BN/KL region, where the surrounding dense envelope of dust and gas conceals 
many of the important processes underway. Mid-infrared observations offer an 
opportunity to probe the physical properties of deeply embedded star formation 
regions. First, mid-infrared wavelengths suffer less extinction than shorter 
wavelengths due to scattering by dust. Second, mid-infrared observations trace 
the presence of warm dust (through thermal emission), which can be used to 
gain an understanding of the overall thermal structure of a region. Third, 
mid-infrared observations can help define the dust properties through the 
detection of spectral features such as the well known silicate band at 
9.7\micron{} and the `unidentified infrared' (UIR) emission bands attributed 
to polycyclic aromatic hydrocarbons, or PAHs. A number of investigators have 
looked at these regions at near- and mid-infrared wavelengths over the past 
30 years: see, \eg{} Becklin \& Neugebauer (1967), Kleinmann \& Low (1967),
McCaughrean \& Gezari (1991), Dougados \etal{} (1993), Gezari \& Backman 
(1994), Hayward, Houck, \& Miles (1994), McCaughrean \& Stauffer (1994), 
Morino \etal{} (1998), Gezari, Backman, \& Werner (1998), and
references therein. See also the chapters by Allen \& Hillenbrand, 
McCaughrean, and Schreyer, Henning, \& Wiesemeyer.

Due to the technology available at the time, most previous mid-infrared 
imaging studies of the region covered either a large area at low spatial 
resolution or a
smaller area in detail. Here we present new large-field, diffraction-limited,
multiwavelength, and deep images covering an approximately 2$\times$2 
arcminute field containing the BN/KL and Trapezium regions in Orion, made
using MIRAC2 (Mid-infrared Array Camera, Hoffmann \etal{} 1998; see also
\hbox{http://cfa--www.harvard.edu/$\sim$jhora/mirac/mirac.html}). These images
have uncovered new information about embedded sources and circumstellar and 
extended material in these regions of active star formation. We present here 
a brief summary of our observational results.

\section{Observations}
Mid-infrared images at 8.8, 10.3, 12.5, and 20.6\micron{} were made using
MIRAC2 on the 3.0\,m NASA Infrared Telescope Facility on 7--11 October 1995.  
Chop-nod images acquired in 3$\times$3 or 4$\times$4 (at 20.6\micron) raster 
patterns centered on the Trapezium were registered using the commanded 
telescope offsets. The image scale at all wavelengths was 0.34 arcsec/pixel. 
The mosaicked field of $\sim$\,2.1$\times$2.1 arcmin covers both the Trapezium 
and BN/KL regions, and the final mosaics are made of 80--170 individual 10 
second images coadded. The mosaics were calibrated using images of the 
standard star $\alpha$\,Tau. The measured FWHMs for $\alpha$\,Tau at 8.8, 
12.5, and 20.6\micron{} were equal to the diffraction limit values of 0.7, 
1.0, and 1.7 arcsec, respectively. The NEFDs (equivalent to the 1$\sigma$ 
detection limit) for a point source are 9, 22, and 200\,mJy, and the 
1$\sigma$ background values are 0.3, 0.5, and 3.0\,mJy, at 8.8, 12.5, and 
20.6\micron{} respectively.

\begin{figure}[t]
\centerline{\psfig{file=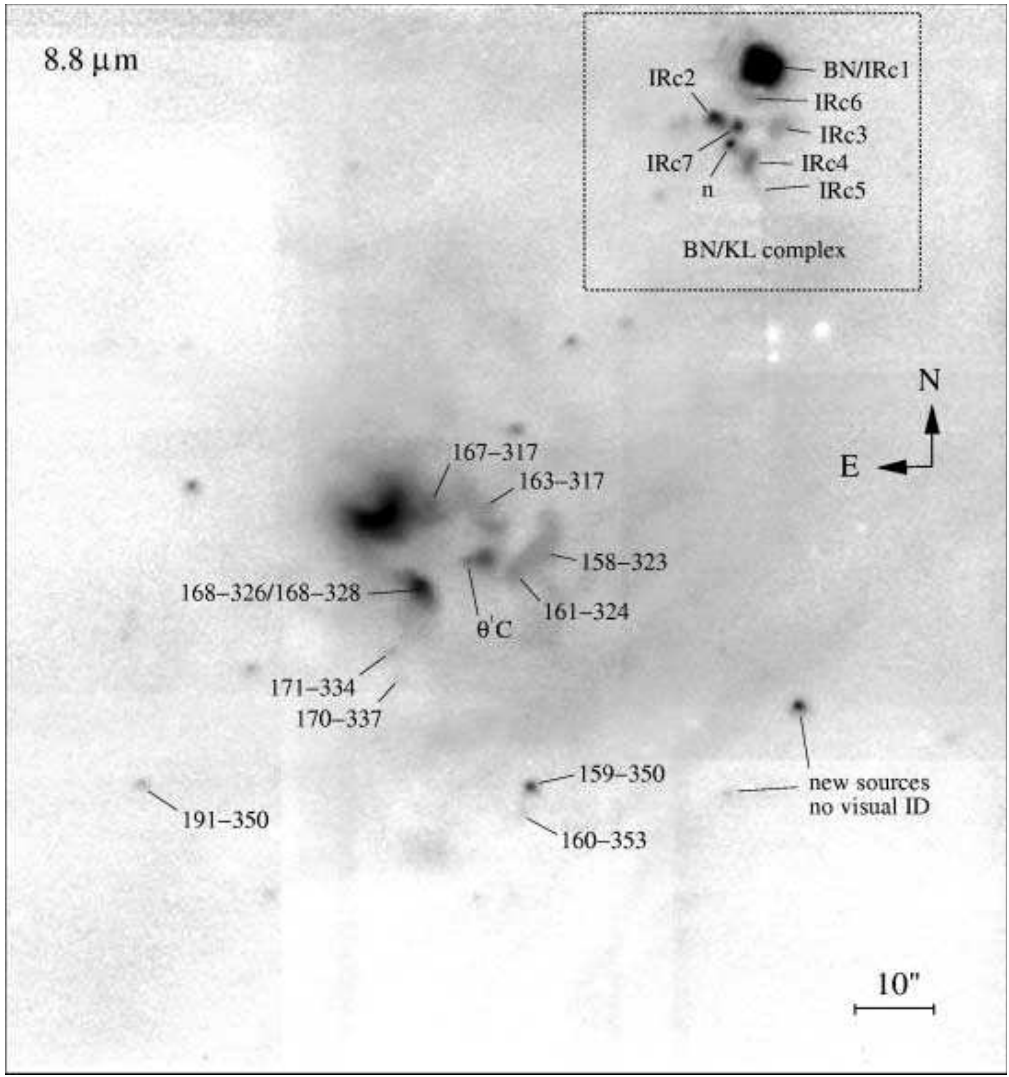,width=\textwidth}}
\caption{8.8\micron{} mosaic of the Trapezium and BN/KL regions covering
$\sim$\,2.1$\times$2.1 arcmin, marked with identifications for various 
sources. The numbers with dashes (\eg{} 191-350) correspond to the locations 
of proplyds previously identified at optical and radio wavelengths. Other 
point sources not specifically identified in this image are stars or other 
known sources which are not proplyds. The scale is shown in the bottom right 
hand side of the mosaic. 
}
\label{fig:eighteight}
\end{figure}

\begin{figure}[t]
\centerline{\psfig{file=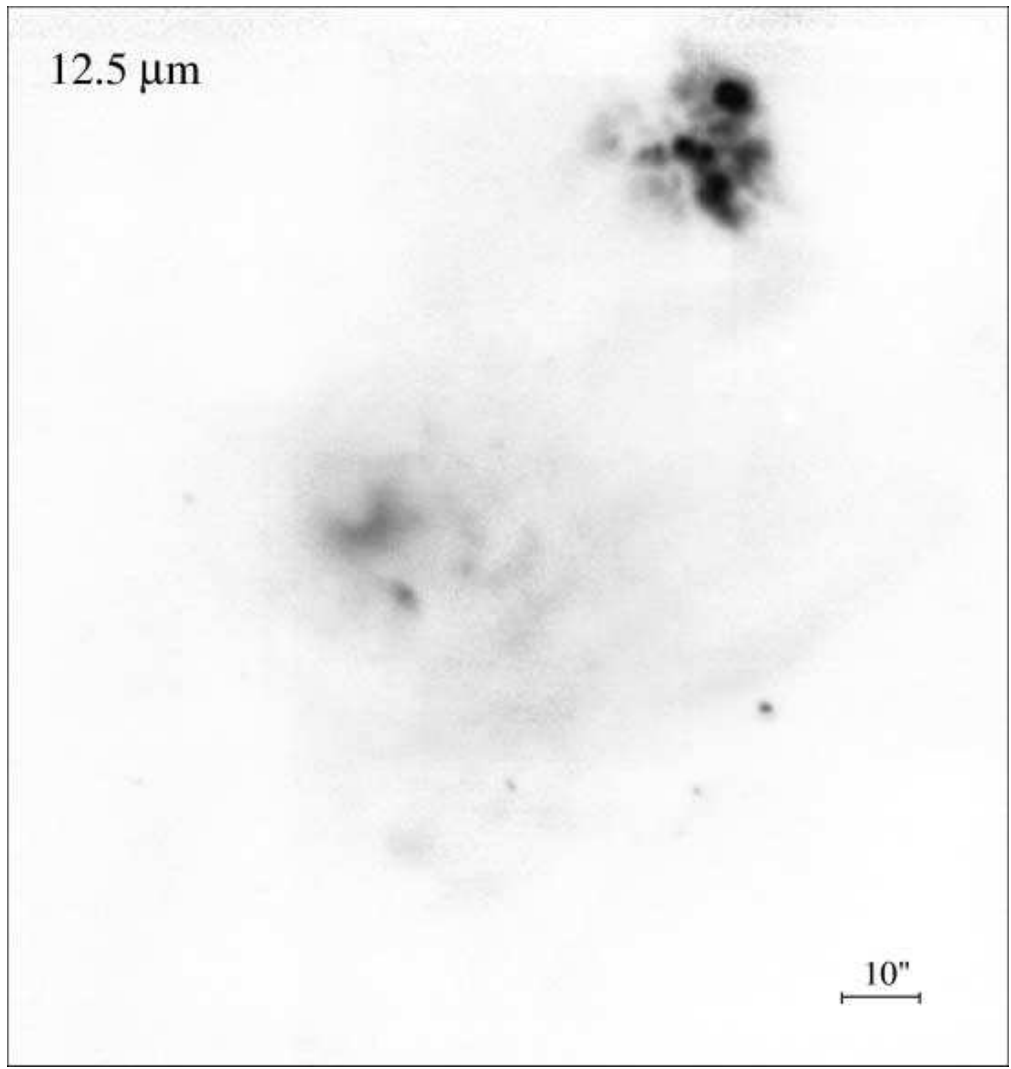,width=\textwidth}}
\caption{12.5\micron{} mosaic of the Trapezium and BN/KL regions covering
roughly the same region as the 8.8\micron{} mosaic in 
Figure~\ref{fig:eighteight}. Note that compared with the 8.8\micron{} mosaic, 
the BN/KL region has become brighter relative to the Trapezium region. An 
extended, clumpy bipolar structure bifurcated by a dark dust lane is seen 
developing at this longer wavelength. A possible jet-like structure can be 
seen extending between IRc3,~4, and~5.} 
\label{fig:twelvefive}
\end{figure}

\begin{figure}[t]
\centerline{\psfig{file=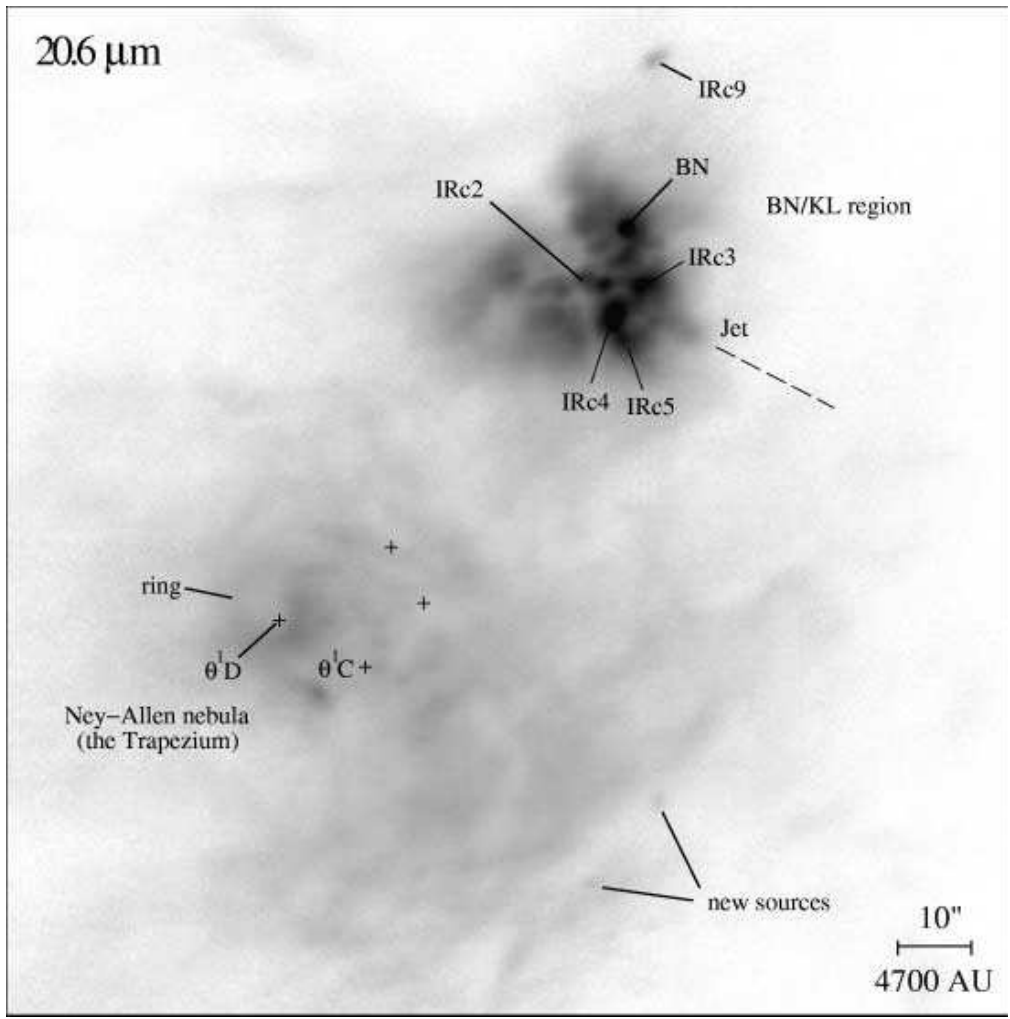,width=\textwidth}}
\caption{20.6\micron{} mosaic of the Trapezium and BN/KL regions. At this 
wavelength, the BN/KL structure clearly dominates. The apparent ring-like 
structure created by IRc3, 4, and~5 can be seen surrounding the jet-like 
extension which points back in the direction of IRc2. The extended `butterfly' 
structure of this region is very pronounced, and IRc9 is now seen to the north 
of the `butterfly'.} 
\label{fig:twentysix}
\end{figure}

\begin{figure}[t]
\centerline{\psfig{file=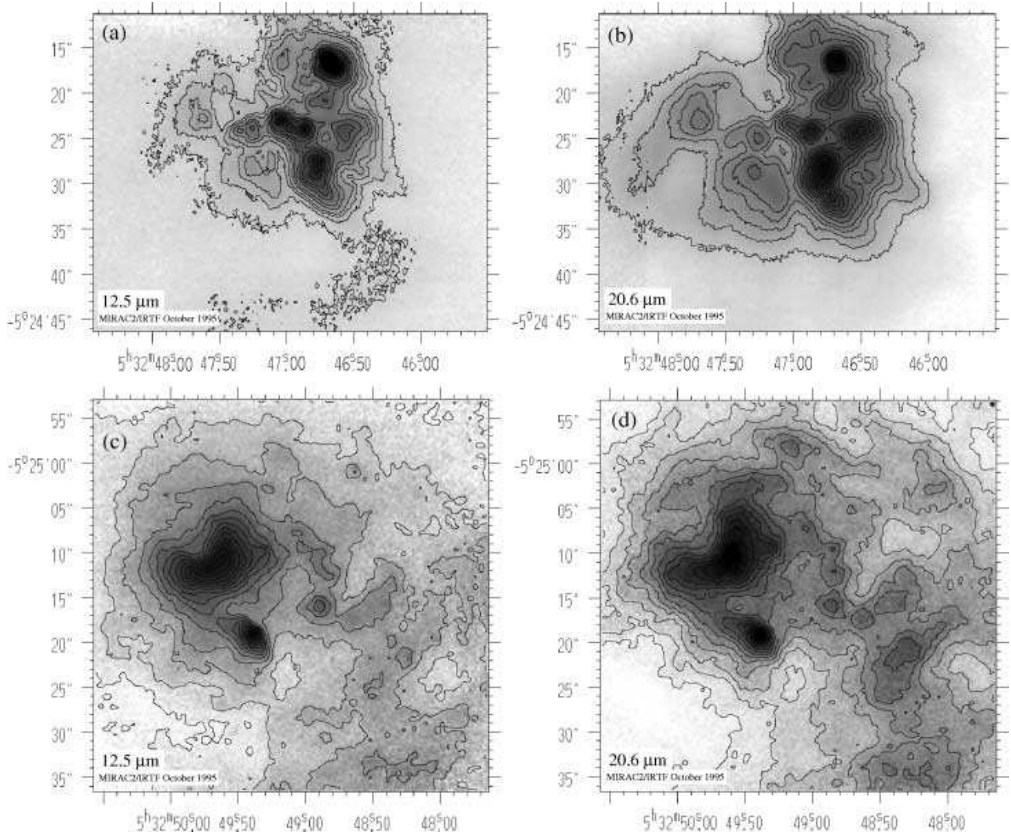,width=\textwidth}}
\caption{Close-up views of the BN/KL (top, panels (a) and (b)) and
Trapezium (bottom, panels (c) and (d)) regions at 12.5 and 20.6\micron{} 
with intensity contours overlaid on the grayscale images.} 
\label{fig:closeup}
\end{figure}

Figures~\ref{fig:eighteight}, \ref{fig:twelvefive}, and~\ref{fig:twentysix}
show the mosaics at 8.8, 12.5, and 20.6\micron{} respectively (the 
10.3\micron{} image is not shown). Figure~\ref{fig:closeup} then shows 
close-up views of the Trapezium and BN/KL regions at 12.5 and 20.6\micron{} 
with overlaid contours.

\section{The Trapezium Region}
In the region surrounding the Trapezium OB stars, we detect mid-infrared 
emission at the locations of eleven proplyds (protoplanetary disks; see
the chapter by McCaughrean) previously identified at optical and radio
wavelengths (Laques \& Vidal 1979; Churchwell \etal{} 1987; Garay, Moran, 
\& Reid 1987; O'Dell, Wen, \& Hu 1993; O'Dell \& Wong 1996). In the 
nomenclature defined by O'Dell \& Wen (1994), these proplyds are
158-323, 159-350, 160-353, 161-324, 163-317, 167-317, 168-326, 168-328, 
170-337, 171-334, and 191-350: their locations are marked on the 8.8\micron{} 
mosaic (Fig~\ref{fig:eighteight}). There are many other proplyds for which 
no mid-infrared emission is detected. The 8.8\micron{} mosaic also shows
previously known arc-like structures (Hayward \etal{} 1994) containing some 
of the proplyds and curving away from \thetaonec.

From our derived 12.5\micron/20.6\micron{} color temperature map (not shown), 
we find that the strongest temperature peak in this region is coincident with
\thetaoned. The extended, diffuse emission in the area which contains most of 
the proplyds around \thetaoned{} is elevated in temperature 
($\sim$\,110--130\,K), and many of the detected proplyds are identified as 
distinct temperature peaks above the level of this diffuse emission. The 
arcs are also distinctly traced as 120\,K structures on top of 110\,K diffuse 
emission.

\begin{sloppypar}
We have also detected two new sources without optical counterparts, about 50 
arcsec southwest of the Trapezium. These lie in the vicinity of the OMC-1S 
complex, a site of ongoing star formation which includes a number of embedded 
near-infrared sources (McCaughrean 1988; Gaume \etal{} 1999), H$_2$O masers 
(Gaume \etal{} 1998), and two or more highly-collimated molecular outflows 
(Ziurys, Wilson, \& Mauersberger 1990; Schmid-Burgk \etal{} 1990; McMullin, 
Mundy, \& Blake 1993; Rodr{\'\i}guez-Franco, Mart{\'\i}n-Pintado, \& Wilson 
1999a, 1999b). Our 12.5\micron/20.6\micron{} color temperature map identifies 
the two sources as strong temperature peaks in the region, suggesting that 
they are self-luminous, with derived color temperatures of 225\,K for the 
brighter northeast source and 180\,K for the fainter 
southwest source. The new sources lie some 10--20 arcsec north of the origin 
of the OMC-1S outflows, and while they have no optical counterparts, they are 
seen to be coincident with very red near-infrared sources on the periphery of 
the OMC-1S core (see the chapter by McCaughrean).
\end{sloppypar}

The Ney-Allen nebula around \thetaoned{} appears to have a ring-like or 
toroidal structure as an extension to its familiar crescent shape. This 
structure is most apparent in the 20.6\micron{} image and color temperature 
maps. In addition, the dust opacity map (not shown) derived from the color
temperature and 12.5\micron{} intensity maps shows an opacity `hole'
coincident with the center of the ring structure.

Finally, the strong, extended, diffuse emission throughout this region
is seen to dominate at the longer wavelengths, with the relative contribution 
more compact sources such as the Ney-Allen nebula and proplyd-containing
structures decreasing with wavelength.

\section{The BN/KL Complex}
The structure of the BN/KL region changes markedly as we look at longer 
mid-infrared
wavelengths. At 8.8\micron{} (Figure~\ref{fig:eighteight}), the BN object or 
IRc1 is by far the brightest source in the field, followed by IRc2, IRc7, and 
source `n' of Lonsdale \etal{} (1982). The individual sources appear relatively 
compact, and there is little apparent extended or diffuse emission associated 
with them. In the 10.3\micron{} image (not shown), silicate absorption is 
clearly present. By 12.5\micron{} (Figure~\ref{fig:twelvefive}), IRc3,~4, 
and~5 are substantially increased in relative brightness compared to the
other sources in the field. A clumpy, apparently bipolar structure is present 
which is bifurcated by a dark dust lane at a position angle of approximately 
30\degree. The brighter half of this `butterfly' includes the known IRc sources 
such as IRc1, 2, 3, 4, 5, and~7, while the fainter half extends to the east 
with some symmetry in the location of clumps or sources between the two sides.
A jet-like structure can be seen in the southwest which appears to lie along 
a line that includes IRc2 and passes through a ring-like structure defined by 
IRc3, 4, and~5. This collimated structure appears to coincide with slow, 
outflowing water masers (Genzel \etal{} 1985; see also the chapter by
Schreyer \etal). At 20.6\micron, the eastern half of the `butterfly' shows 
even greater extension and intensity, and the symmetry of the structures in
the two halves is quite pronounced. At both 12.5 and 20.6\micron, a faint 
source appears just to the northeast of IRc2 in the center of a large, dark 
`hole' between the two sides at the top of the dark lane. This source may lie 
along the line through the jet-like structure to the southwest. Also, by 
20.6\micron, IRc4 becomes brighter than the BN object, and IRc9 can be seen 
as an elongated object north of BN (only the 20.6\micron{} image extends far 
enough north to include IRc9).

A derived color temperature map reveals that BN and IRc2 are 
the dominant temperature peaks in the region, in agreement with previous 
results (Wynn-Williams \etal{} 1984; see also the chapter by Schreyer \etal). 
Temperatures in the region away from these two temperature peaks fall in 
the range of 100--140\,K\@. IRc3 and~5 do not correspond to color temperature 
peaks, but do correspond to dust opacity peaks, suggesting that 
they are not self-luminous sources. IRc4 and the faint source northeast of 
IRc2 correspond to slight local temperature increases. However, IRc4 is a 
definite opacity peak, while the faint northeast source is in an opacity hole. 
The pattern of the temperature contours follows the ring-like structure seen 
in the intensity images for IRc3, 4, and~5.  Another source in the eastern 
half of the `butterfly', just to the east of IRc2, shows up as both a local 
temperature peak and an opacity minimum. Finally, we see that the entire 
BN/KL region has a much higher average opacity than the region around the
Trapezium OB stars.

This brief overview of the results from our imaging work will be followed
by a full analysis in another paper presently in preparation.

\acknowledgments

The MIRAC2 project has been supported by the National Science Foundation, 
the National Aeronautics and Space Administration, Steward Observatory of 
the University of Arizona, and the Smithsonian Astrophysical Observatory.  
Lynne Deutsch and William Hoffmann were visiting astronomers at the NASA 
Infrared Telescope Facility. We thank Mrinal Iyengar, Mari Paz Miralles, and
the IRTF staff for assistance at the telescope.

\end{document}